# Effects of metals (X = Pd, Ag, Cd ) on structural, electronic, mechanical, thermoelectric and hydrogen storage properties of LiXH$_3$ perovskites

Anupam[a], Shyam Lal Gupta[b], Sumit Kumar[c], Samjeet Singh Thakur[d], Sanjay Panwar[a,*] and Diwaker[e,**]

[a]*School of Basic and Applied Sciences, Maharaja Agrasen University, Baddi, Solan, 174103, H P, INDIA*
[b]*Department of Physics, HarishChandra Research Institute, Prayagraj, Allahabad, 211019, U P, INDIA*
[c]*Department of Physics, Government College, Una, 174303, H P, INDIA*
[d]*Department of Chemistry, NSCBM Government College, Hamirpur, 177005, H P, INDIA*
[e]*Department of Physics, SCVB Government College, Palampur, Kangra, 176061, H P, INDIA*

**ABSTRACT**

Using the WIEN2K code, the hydrogen storage capabilities of lithium compositions like LiXH$_3$ (X = Pd, Ag, Cd) hydrides are examined. Structural, electrical, mechanical, thermoelectric, and hydrogen storage properties of these hydrides are analyzed using first-principles simulations to verify their stability. Structural analysis of these compositions reveals that the hydrides are stable and belong to the cubic space group number (221 Pm-3m). The thermodynamic stability of these hydrides are given in term of gravimetric hydrogen storage capacities. The purpose of the study is to calculate calefaction of formation and breakdown temperature to determine stability of these hydrides. The metallic nature of all compositions are confirmed by band plots and density of states. The elastic properties such as elastic constant, Pugh's ratio, bulk modulus, Poisson's ratio and anisotropy factor are calculated to check the applicability of these compositions for applications involving hydrogen storage. The present paper represents the initial theoretical approach toward the future exploration of these material for hydrogen storage applications.

## 1. Introduction

Energy crises are a result of the rising demand for fossil fuels like coal, oil, and petroleum. Consuming fossil fuels causes the release of dangerous toxic gases like chlorofluorocarbon (CFC), $CO_2$, $N_2O$, $CH_4$, and others that harm the environment. As a result, it is challenging for researchers around the world to look into other alternative sources of energy(Anupam, Gupta, Kumar, Panwar and Diwaker, 2024; Al, Cavdar and Arikan, 2024; Ullah, Riaz and Rizwan, 2023; Mera and Rehman, 2024; Ahmed, Tahir, Ali and Sagir, 2024a; Mbonu, Louis, Chukwu, Agwamba, Ghotekar and Adeyinka, 2024; Song, Chen, Chen, Xu and Zhang, 2024; Azeem, Shahzad, Wong and Tahir, 2024; Xu, Chen, Chen, Li and Zhang, 2024). Since hydrogen is a clean, non-toxic, and renewable source of energy, it is used as a substitute for existing non-renewable energy sources to fulfill the need for energy. Hydrogen is abundant in nature and can be used for a variety of purposes, including cooking, electricity production, running factories and aero planes, the automobile industry, and supplying domestic energy needs. With hydrogen-powered cars, the automobile industry has a lot of potential in the transportation sector. Significant obstacles exist for both the extraction and storage of hydrogen

(Ahmed, Tahir, Ali and Sagir, 2024b; Bahhar, Tahiri, Jabar, Louzazni, Idiri and Bioud, 2024; Ghani, Sagir, Tahir, Ullah and Assiri, 2024; Siddique, Khalil, Almutairi, Tahir, Sagir, Ullah, Hannan, Ali, Alrobei and Alzaid, 2023; Rkhis, Laasri, Touhtouh, Hlil, Bououdina, Ahuja, Zaidat, Obbade and Hajjaji, 2022; Raza, Murtaza, e Hani, Muhammad and Ramay, 2020; Khalil, Hussain, Hussain, Rana, Murtaza, Shakeel and Asif Javed, 2021; Al, Yortanlı and Mete, 2020; Bellosta von Colbe, Ares, Barale, Baricco, Buckley, Capurso, Gallandat, Grant, Guzik, Jacob, Jensen, Jensen, Jepsen, Klassen, Lototskyy, Manickam, Montone, Puszkiel, Sartori, Sheppard, Stuart, Walker, Webb, Yang, Yartys, Züttel and Dornheim, 2019; Padhee, Roy and Pati, 2022; Padhee, S. P., Roy, A. and Pati, S., 2021). Its storage density needs to be increased to make it commercially feasible. For the storage of hydrogen at high density, several techniques have been developed, including cryogenic form, compound form, and pressurized gas form. The second method for storing hydrogen is the most promising, aspirational, and developing. The storage of hydrogen in solids is preferred over that of liquid and pressurized gas. Metal organic frameorks and carbon nanotube can store hydrogen using the chemisorption or physisorption processes. The high volumetric density of hydrogen atoms in the host lattice makes metal hydrides suitable for hydrogen storage applications as well. Favorable thermodynamic characteristics and structural stability of metal hydrides are additional factors that make them appropriate for applications involving hydrogen storage (Chattaraj, D., Dash, S. and Majumder, C., 2016). Researchers are currently facing a significant challenge in fully comprehending the kinds of materials that can deposit large amounts of hydrogen as fuel(Zapatero, P. A., Herrero, A., Lebon, A, Gallego, L. J. and Vega,A., 2021; Abdellaoui, M., Lakhal, M., Benzidi, H, Mounkachi, O, Benyoussef, A., Kenz, A. E. and Loulidi, M., 2020). In our efforts towards that we have recently studied Zirconium based ZrXH₃ (X = Zn, Cd) hydride perovskites and found them to be of reasonable importance for hydrogen storage applications(Anupam et al., 2024). In continuation to that we are now presenting another study on Li based compositions LiXH₃ (X=Pd, Ag, Cd) perovskite using DFT and other related methodologies as implemented in WIEN2K(Blaha, P., Schwarz, K., Tran, F., Laskowski, R., Madsen, G. K. H. and Marks, L. D., 2020). to explore the structural, mechanical, electronic, optical and thermal properties along with their hydrogen storage capabilities.

*Corresponding author
**Principal corresponding author
dr.spanwar@gmail.com (S. Panwar); diwakerphysics@gmail.com ( Diwaker)
ORCID(s): 0000-0003-3283-3165 (S. Panwar); 0000-0002-4155-7417 ( Diwaker)

## 2. Method of Calculations

We examined the properties of LiXH₃ (X=Pd, Ag, Cd) perovskite using the full-potential linearized augmented plane wave method as implemented in WEIN2K code(Blaha, P. et al., 2020). Throughout, we have used the Perdew Burke Ernzerhof-generalized gradient approximation exchange correlation potential to investigate various properties of theses compositions characteristics. Using Birch-Murnaghan's equation of state, we were able to obtain the energy versus volume curve and determine the stable phase and structure parameter of these hyrides. The elastic constant for our compositions is determined using the IRElast package as integrated with WIEN2K code. Our calculations use a default cutoff energy value of -6.0 Ryd. Throughout the calculations, the RMT value and k points are set to be 7.0 and 1000, respectively.

## 3. Results and Discussions

In this section we will discuss the structural, electronic, mechanical and thermoelectric properties of lithium compositions LiXH₃with (X=Pg, Ag, Cd).

### 3.1. Investigations of Structural Phase Stability on LiXH₃with (X=Pg, Ag, Cd)

The perovskites of the LiXH₃ (X=Pd, Ag, and Cd) hydrides have a cubic crystal structure and belong to space group 221 Pm-3m. These combinations have undergone optimization and the optimized structures are presented in figure 1 as shown below. Figure 2 shows plot of Birch-Murnaghan's equation of state for energy vs volume of each combination. In the cubic unit cell as represented in figure 1 transition metals, lithium atoms and hydrogen atoms are seated at body centered, corners and face centered positions respectively. The Wyckoff positions of X, Li are (0.5, 0.5, 0.5) and (0, 0, 0) respectively where as for hydrogens $H_1$, $H_2$ and $H_3$ are (0, 0.5, 0.5), (0.5, 0, 0.5) and (0.5, 0.5, 0). Using Birch-Murnaghan's equation of states the computed lattice constant for these compositions comes out to be 3.5, 3.6 and 3.9 Å for LiXH₃ (X=Pd, Ag, Cd), respectively. Other parameters which has been computed from Birch-Murnaghan's equation are volume ($V_0$), ground state energy ($E_0$), bulk modulus (B) and pressure derivative of bulk modulus (B′) respectively presented in Table 1.

### 3.2. Formation enthalpies and gravimetric capacities

Two important parameters required to evaluate the thermodynamic properties of inter metallic hydrides are decomposition temperature ($T_d$) and enthalpy of formation ($\Delta H_f$). We have calculated these parameters our compositions which will support the dehydrogenation characteristics and phase constancy of LiXH₃with (X=Pg, Ag, Cd) hydrides perovskite .The best value for hydrogen storage material for hydrides is about -40kJ/mol.H₂ as specified by US department of energy[27]. The underlying mechanism will be explained using the following equations that leads to the formation of LiXH₃with (X=Pg, Ag, Cd)hydrides perovskite(Sato, Noréus, Takeshita and Häussermann, 2005).

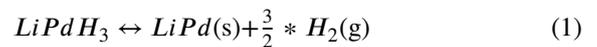

$$LiPdH_3 \leftrightarrow LiPd(s) + \tfrac{3}{2} * H_2(g) \qquad (1)$$

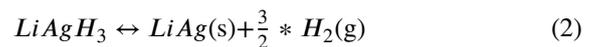

$$LiAgH_3 \leftrightarrow LiAg(s) + \tfrac{3}{2} * H_2(g) \qquad (2)$$

and

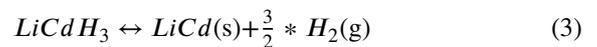

$$LiCdH_3 \leftrightarrow LiCd(s) + \tfrac{3}{2} * H_2(g) \qquad (3)$$

Using Hess law as given below

$$\Delta H_f = \sum E_{tot}(\text{products}) - \sum E_{tot}(\text{reactants}) \qquad (4)$$

The enthalpy of formation for these compositions with (X=Pg, Ag, Cd) is calculated using the equation given below

$$\Delta H_f = E_{tot}(LiXH_3) - E_{tot}(LiX) - \tfrac{3}{2}E_{tot}(H_2) \qquad (5)$$

The estimated enthalpy of formation for these compounds comes out to be -55.05 kJ/mol.H₂ for LiPdH₃, -24.36 kJ/mol.H₂ for LiAgH₃ and -49.71 kJ/mol.H₂ for LiCdH₃, correspondingly, with the FP-LAPW approach, which is higher than the optimal value of -40 kJ/mol.H₂ (US energy department)(Klebanoff, L.E. and Keller, J. O., 2013). This show that these perovskite have high thermodynamic stability. Further, the decomposition temperature of these perovskite is calculated by the relation of Gibbs energy $\Delta G$ given as

$$\Delta G = \Delta H - T\Delta S \qquad (6)$$

where $\Delta H$ and $\Delta S$ stand for the formation enthalpy and entropy change of the dehydrogenation process, respectively. At equilibrium, the standard value of Gibbs energy is zero i.e ($\Delta G$=0). The development of hydrogen molecules is the primary source of the dehydrogenation reaction's change in entropy, which is extremely important. This change in entropy is considered for intermetallic hydrides as ($\Delta S = 130$ J/mol.K$^{-1}$(Zeng, Q., Su, K., Zhang, L., Xu, Y., Cheng, L. and Yan, X., 2006). So the decomposition temperature is reported as

$$T_d = \frac{|\Delta H|}{130} \qquad (7)$$

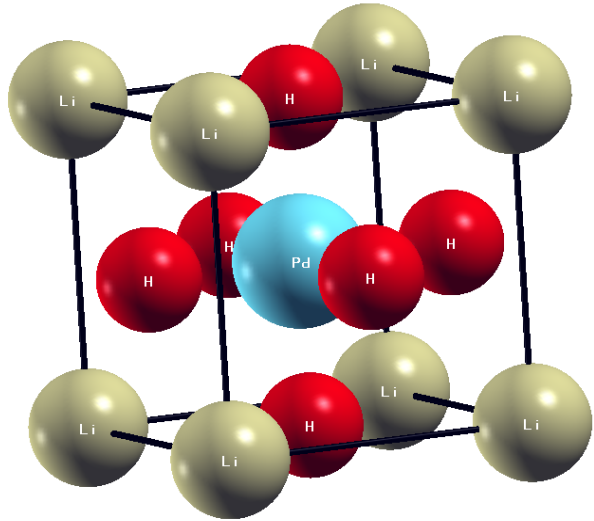

Figure 1: Optimised crystal structure of LiXH$_3$ with (X=Pd, Ag, Cd)

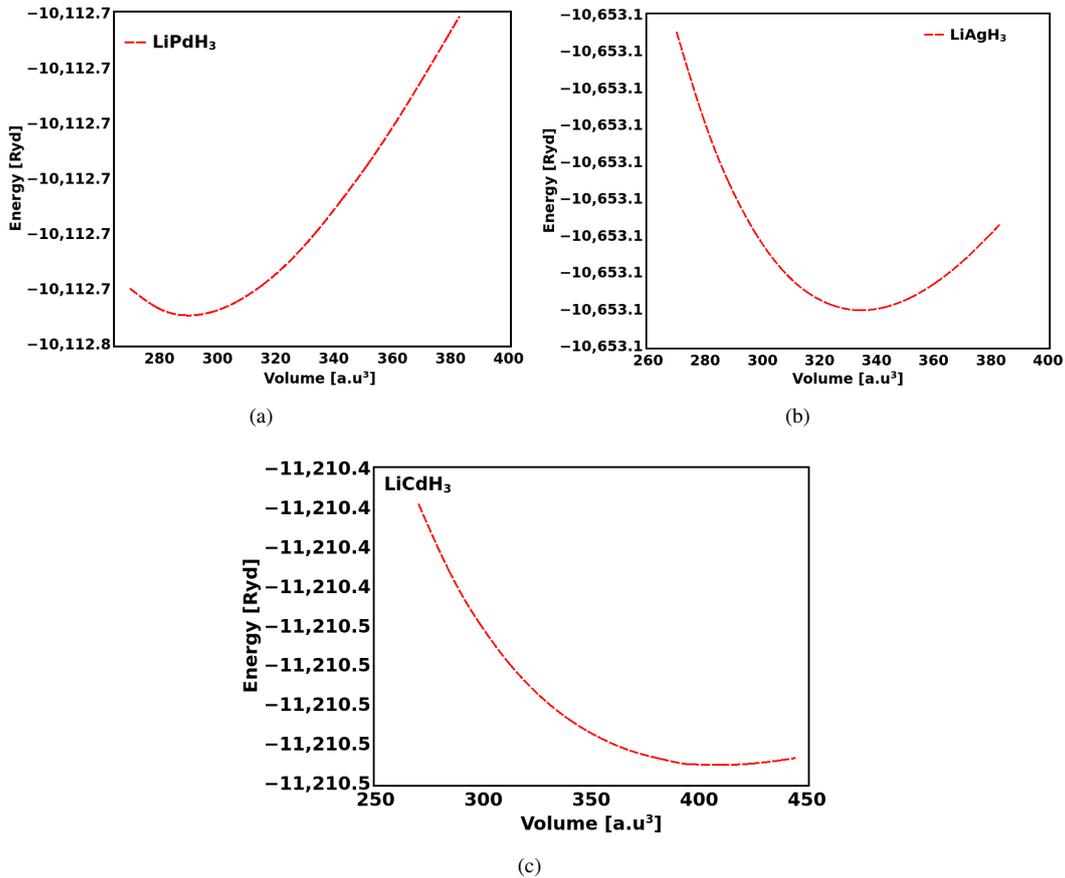

Figure 2: Energy versus Volume curves of (a) LiPdH$_3$ (b) LiAgH$_3$ (c) LiCdH$_3$

By replacing the enthalpy of formation value from Table 1 with the calculated decomposition temperature value in the preceding equation for LiXH$_3$ with (X=Pg, Ag, Cd) hydrides type perovskites comes out to be 423 K for LiPdH$_3$, 187 K for LiAgH$_3$ and 382 K for LiCdH$_3$ respectively, which confirm the higher stability of these compositions. Furthermore, it's critical to determine the compositions' gravimetric hydrogen storage capacity in order to take into account the suitability of these hydrides for use in hydrogen storage applications.(Ikeda, K., Kato, S., Shinzato, Y., Okuda, N., Nakamori, Y., Kitano, A., Yukawa, H., Morinaga, M. and Orimo, S., 2007). The following formula determines the gravimetric hydrogen storage capacity, which is the amount of hydrogen stored per unit mass of a substance.(Broom, D. P., Webb, C. J., Fanourgakis, G. S., Trikalitis, P. N. and Hirscher, M., 2019; Züttel, A., 2003).

Table 1
Lattice parameter [a (Å)], Bulk Modulus ( B in GPa), Volume ($V_0$ in Å$^3$), Pressure derivative of Bulk modulus ( B$'$ in GPa), Ground state Energy ($E_0$ (eV)), enthalpy of formation ($\Delta H_f$ (KJ/mol.H$_2$)) and Gravimetric storage densities ($C_{wt\%}$), tolerance factor (t) and octahedral factor ($\mu$).of LiXH$_3$with (X=Pg, Ag, Cd) perovskites.

| Perovskite | a | B | $V_0$ | B$'$ | $E_0$ | $\Delta H_f$ | $C_{wt\%}$ | t | $\mu$ |
|---|---|---|---|---|---|---|---|---|---|
| LiPdH$_3$ | 3.5 | 86.8 | 290.3 | 4.13 | -10112.74 | -55.05 | 2.7 | 0.7 | 0.6 |
| LiAgH$_3$ | 3.6 | 61.4 | 334.5 | 4.20 | -10653.13 | -24.36 | 2.5 | 0.7 | 0.7 |
| LiCdH$_3$ | 3.9 | 40.7 | 410.4 | 3.98 | -11210.53 | -49.71 | 2.4 | 0.7 | 0.7 |

$$C_{wt\%} = \left[\frac{n_H * m_H}{M_{(LiXH_3)}} \times 100\right]\% \qquad (8)$$

In this case, the number of hydrogen atoms is $n_H$, the molar mass of hydrogen is $m_H$, and the molar mass of the compound is $M_{(LiXH_3)}$. From the above formula after calculation, the gravimetric hydrogen storage capacity is found to be 2.71%, 2.57% and 2.47% for LiXH$_3$ (X=Pd,Ag,Cd) respectively. After comparison it has been found that LiPdH$_3$ has highest value i.e (2.71%) indicating that among three specified composition LiPdH$_3$ have high gravimetric storage capacity. In comparison to their corresponding bulk compositions, further thin films and nanostructures with a significant surface area would have improved hydrogen storage capabilities. (Park, S. J. and Seo, M. K., 2011). Moreover, Goldschmidt factor (t) and octahedral factor ($\mu$) are also computed to assess the structural stability of these compositions. (Khan, U A., Khan, N. U., Abdullah, Alghtani, A. H. , Tirth, V, Ahmed, S. J, Sajjad, M., Alghtani, A., Shaheed, T. and Zaman, A., 2022) using the equations given below

$$t = \frac{r_{Li} + r_H}{\sqrt{2}[r_{(X=Pd,Ag,Cd)} + r_H]} \qquad (9)$$

and

$$\mu = \frac{r_X}{r_H} \qquad (10)$$

where $r_{Li}$ and $r_H$ are the radii of Li and Hydrogen while $r_X$ is the radii of (X=Pd, Ag, Cd) respectively. For the cubic perovskite structures this value of t should be in the range 0.7-1 and $\mu$ should be in the range 0.4-0.7. The calculated values of t and $\mu$ as reported in Table 1 are in the above mentioned range and indicate that all these compositions exhibit cubic symmetric crystal structure that is stable. The enhanced stability of hydrogenated compositions compared to respective pristine matrices and low value of decomposition temperature indicating the enhanced hydrogen reversibility in these compositions

### 3.3. Electronic Structure

To determine the electronic structure of lithium-based perovskite i.e. LiXH$_3$ (X=Pd,Ag,Cd) energy band gap, density of states, and partial density of states have been evaluated. The energy band gap is formed by valence band maximum and conduction band minimum(Collins, P. G., Bradley, K., Ishigami, M. and Zettl, A., 2000). It is quite clear from the Figures 3-5 (a) the metallic nature of LiXH$_3$ (X=Pd,Ag,Cd) as confirmed by the fair overlapping of the conduction band and valence band. Figures 3-5 (b) shows total or partial density of states in LiXH$_3$ (X=Pd,Ag,Cd) perovskites. Furthermore, the total and partial density of states of LiXH$_3$ (X=Pd,Ag,Cd) perovskites allow for a clear prediction of the atoms' contribution to the band structures. In LiXH$_3$ (X=Pd,Ag,Cd) perovskites, X elements are found to be most productive contributors towards partial density of states. Figures 3-5(b) clearly indicates that the s and p states of Li-atom where as d-states of X ( X = Pd, Ag, Cd) contributes more to conduction band however, s and p states of X (X = Pd, Ag, Cd) have less impact on band structure. Also, s states of hydrogen atom seem to contribute in both conduction band and valence band as indicated fro the total and partial density of states. The d states of transition metal (X = Pd, Ag, Cd) of these compositions are crossing the fermi-energy ($E_F$) level which states that these composition are metallic in nature. These metallic hydrides(Bowman, R. C. and Fultz, B., 2002) LiXH$_3$ have broad area of applications in hydrogen storage materials such as rechargeable batteries, automotive fuel storage, jet planes as fuel and to run factories etc.

### 3.4. Mechanical Properties

Since the elastic constants ($C_{ij}$) that describe the material's response to applied stresses rely on the crystal structure, hence obtaining the elastic constants is very essential in order to assess a material's mechanical stability, which is crucial for technological applications(Anupam., Gupta, S. L., Kumar, S., Panwar, S. and Diwaker, 2024; Surucu, G., Gencer, A., Candan, A., Gullu, Hasan H. and Isik, M., 2020; Anupam et al., 2024). Table 2 provides information on the key elastic constants for the cubic perovskites LiXH$_3$ (X=Pd,Ag,Cd) i. e. C$_{11}$, C$_{22}$, and C$_{44}$. These elastic constants are mechanically stable because they satisfy the specified Born stability conditions, which are C$_{11}$ > 0, C$_{44}$ > 0, (C$_{11}$-C$_{12}$)>0, (C$_{11}$+2C$_{12}$)>0. For compressive force (pressure) of 0 GPa, 5 GPa, and 10 GPa, the model integrated with the most recent version of WIEN2k and implemented in the IRelast package (Wang, J., Sidney, Y., Phillpot, S. R. and Dieter, W., 1993) is used to compute these elastic constants. To determine the toughness and stability of a compound, these constants are computed at a certain pressure by calculating the stress tensor components for little deformation at constant volume(Xu, Y., Peng, B., Zhang, H., Shao, H., Zhang, R. and Zhu, H., 2018) . Diamond Anvil Cells may be used in experiments to obtain them(Levitas, V. I., Kamrani, M. and Feng, B., 2019). Certain physical characteristics, such bulk modulus and shear modulus, could be determined using the computed elastic constants. By considering the Voigt-Reuss-Hill equation's theoretical approximation, different mechanical elastic parameters are calculated for LiXH$_3$ (X=Pd,Ag,Cd) perovskite under external stress of 0, 5 and 10 GPa (Razzaq, S., Murtaza, G., Khalil, R. M. A., Ahmad, N. and Raza, H. H., 2022) and are tabulated in Table 2. The different kind of bulk modulus is calculated by using the values of elastic constants and is given as

$$B = \frac{1}{3}\left[c_{11} + 2c_{12}\right] = \frac{1}{3}\frac{1}{(s_{11} + 2s_{12})} = B_V = B_R \quad (11)$$

where

$$s_{11} = \frac{c_{11} + c_{12}}{c}; s_{12} = \frac{-c_{12}}{c}; \quad (12)$$

and

$$c = (c_{11} - c_{12})(c_{11} + 2c_{12}); s_{44} = \frac{1}{c_{44}} \quad (13)$$

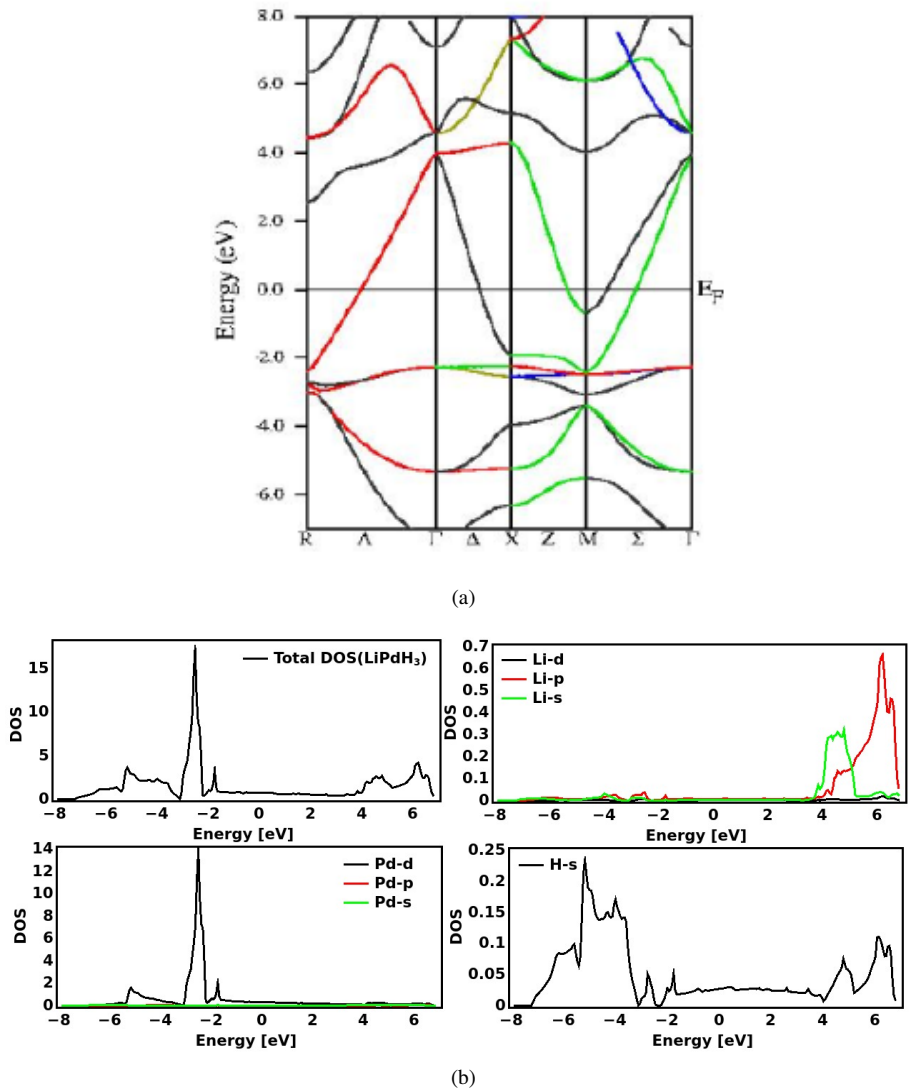

Figure 3: [a] Energy Band Structures and [b] Density/Partial density of states of LiPdH$_3$

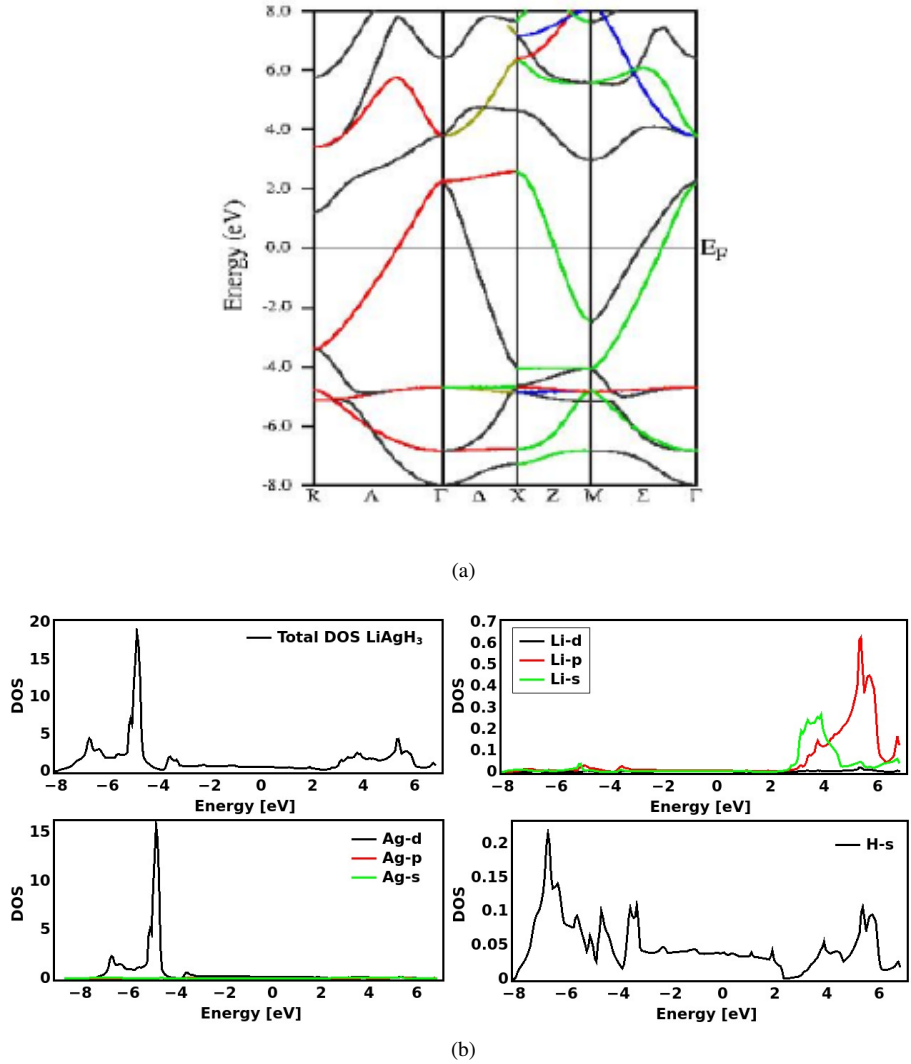

(a)

(b)

**Figure 4**: [a] Energy Band Structures and [b] Density/Partial density of states of LiAgH$_3$

The cubic crystal's Cauchy pressure (CP), Pugh's ratio (k), and Poisson's ratio ($v$) are determined using the elastic constants given by

$$CP = (c_{12} - c_{44}); k = \frac{B}{G}; G = \frac{c_{11} - c_{12} + 3c_{44}}{5} \quad (14)$$

and

$$v = \frac{c_{12}}{c_{11} + c_{12}} \quad (15)$$

The Voigt's ($G_V$) and Reuss ($G_R$) shear modulus are also computed from elastic constants and are given as

$$G_V = \frac{1}{5}(c_{11} - c_{12} + 3c_{44}) \quad (16)$$

and

$$G_R = 5\left[\frac{(c_{11} - c_{12})c_{44}}{3(c_{11} - c_{12}) + 4c_{44}}\right] \quad (17)$$

The shear modulus (G) and Poisson's ratio ($v$) are used to calculate the Young's modulus given as

$$E = 2G(1 + v) \quad (18)$$

Ranganathan-Anisotropy Index ($A_u$) for cubic crystals is computed from the equation as

$$A_u = 5\left[\frac{G_v}{G_R} - 1\right] \quad (19)$$

Kleinman's parameter ($\zeta$) for our compositions is computed as per equation given as

$$\zeta = \frac{c_{11} + 8c_{12}}{7c_{11} + 2c_{12}} \quad (20)$$

It shows how atoms in various sub lattices are positioned in relation to one another when volume-conserving strain distortions are present and the symmetry no longer holds the positions in place. The volume of the crystal affects the Debye temperature $\theta_D$. $\theta_D$ is exactly determined for each volume V in terms of the elastic constants through a spherical average of the three components of the sound velocity. It is provided as

$$\theta_D = \frac{\hbar}{k_B} 3\sqrt{n \ast 6\pi^2 \sqrt{V}} \sqrt{\frac{B}{M} f_\nu} \qquad (21)$$

where $k_B$ is the Boltzmann constant, $\hbar$ is Planck's constant, and n is the number of atoms in the primitive cell with volume V unit. A function of the Poisson ratio $\nu$, denoted as $f_\nu$, and M, the compound's mass, correspond to V.

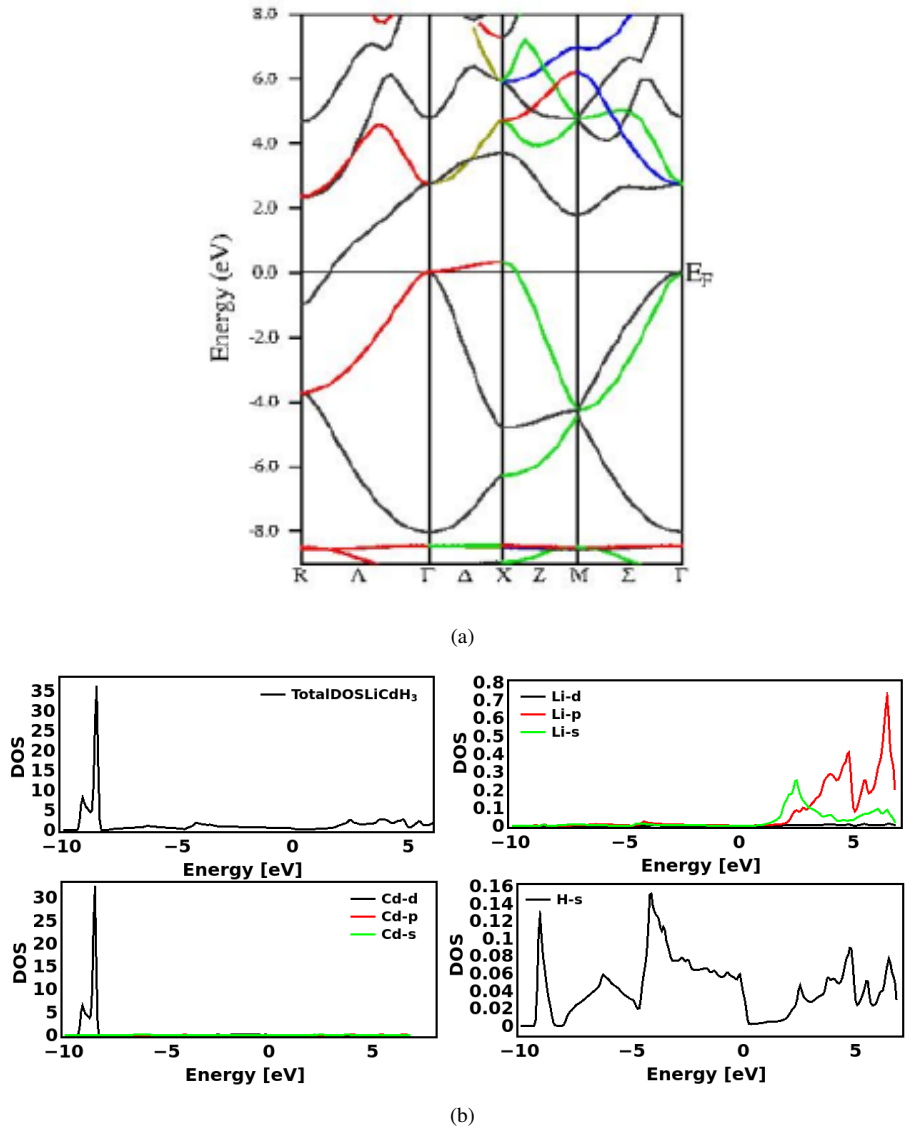

Figure 5: [a] Energy Band Structures and [b] Density/Partial density of states of LiCdH$_3$

$$f_\nu = 3\sqrt{\frac{3}{2\left[\frac{2(1+\nu)}{3(1-2\nu)}\right]^{\frac{3}{2}} + \left[\frac{(1+\nu)}{3(1-\nu)}\right]^{\frac{3}{2}}}} \qquad (22)$$

Another important mechanical parameter is the degree of hardness of a material. It is not possible to calculate hardness using only the elastic constants. To relate the elastic moduli of a material to its hardness, several models has been put forward. Pugh used the formula $H_B = G\frac{b}{c}$, where b is the dislocation's Burger vector and c is a constant for all metals with the same structural composition, to create a link between the shear modulus (G) and the Brinell hardness ($H_B$) of pure metals. Teter discovered the semi-empirical link between Vicker's hardness $H_V$ and the rigidity modulus (G) as $H^{T_V} \sim 0.151G$. Additionally, Chen et al. provided a semi-empirical relationship between the shear modulus (G) and the squared Pugh's ratio ($k = \frac{B}{G}$), which determines Vicker's hardness.

$$H^{C_V} = 2(k^{-2}G)^{0.585} - 3 \qquad (23)$$

**Table 2**
Elastic constants (C$_{ij}$), Bulk, Shear and Young Modulus (B$_V$, B$_R$, B, G$_V$, G$_R$, G and E$_V$, E$_R$, E in GPa), Reuss and Hill Poisson's coefficient($\nu_V$, $\nu_R$ in GPa), Kleinman's parameter ($\zeta$), Ranganathan and Kube Anisotropy Index (A$_u$, A$_k$), Transverse, Longitudnal and Average wave velocity (V$_t$, V$_l$ and V$_a$ in m/s), Debye Temperature ($\theta_D$ in K), Pugh's Ratio (k), Chen and Tian Vickers hardness ($H^{C_V}$, $H^{T_V}$ in GPa), Lame's first and second parameter ($\lambda$, $\mu$ in GPa) of LiXH$_3$ (X=Pd,Ag,Cd) perovskite under external stress of 0, 5 and 10 GPa.

| Stress | 0 GPa | | | 5 GPa | | | 10 GPa | | |
|---|---|---|---|---|---|---|---|---|---|
| Parameters | LiPdH$_3$ | LiAgH$_3$ | LiCdH$_3$ | LiPdH$_3$ | LiAgH$_3$ | LiCdH$_3$ | LiPdH$_3$ | LiAgH$_3$ | LiCdH$_3$ |
| C$_{11}$ | 267.718 | 121.328 | 44.271 | 261.058 | 163.887 | 90.094 | 303.282 | 230.544 | 128.207 |
| C$_{12}$ | 85.666 | 31.187 | 29.097 | 30.442 | 35.831 | 42.566 | 34.917 | 31.308 | 47.519 |
| C$_{11}$-C$_{12}$ | 182.052 | 90.141 | 15.174 | 230.616 | 128.056 | 47.528 | 268.365 | 199.236 | 80.688 |
| C$_{11}$+2C$_{12}$ | 439.050 | 183.704 | 102.467 | 321.943 | 235.549 | 175.226 | 373.117 | 293.162 | 223.246 |
| C$_{44}$ | 21.672 | 10.972 | 10.308 | 26.753 | 15.700 | 13.479 | 34.492 | 23.803 | 13.189 |
| $CP = C_{12}$-C$_{44}$ | 246.046 | 20.215 | 18.789 | 3.689 | 20.131 | 29.087 | 268.79 | 7.505 | 34.33 |
| B$_V$ | 146.350 | 61.234 | 34.155 | 107.314 | 78.516 | 58.409 | 124.372 | 97.720 | 74.415 |
| B$_R$ | 146.350 | 61.234 | 34.155 | 107.314 | 78.516 | 58.409 | 124.372 | 97.720 | 74.415 |
| B | 146.350 | 61.234 | 34.155 | 107.314 | 78.516 | 58.409 | 124.372 | 97.720 | 74.415 |
| G$_V$ | 49.414 | 24.611 | 9.220 | 62.175 | 35.031 | 17.593 | 74.368 | 54.129 | 24.051 |
| G$_R$ | 31.172 | 15.733 | 9.015 | 38.615 | 22.490 | 16.301 | 49.076 | 34.221 | 18.048 |
| G | 40.293 | 20.172 | 9.117 | 50.395 | 28.761 | 16.947 | 61.722 | 44.175 | 21.050 |
| E$_V$ | 133.245 | 65.111 | 25.376 | 156.333 | 91.487 | 47.963 | 186.026 | 137.077 | 65.136 |
| E$_R$ | 87.317 | 43.476 | 24.857 | 103.439 | 61.590 | 44.741 | 130.115 | 91.930 | 50.708 |
| E | 110.718 | 54.529 | 25.117 | 130.723 | 76.893 | 46.358 | 158.884 | 115.170 | 57.708 |
| $\nu_V$ | 0.348 | 0.323 | 0.376 | 0.257 | 0.306 | 0.363 | 0.251 | 0.266 | 0.354 |
| $\nu_R$ | 0.401 | 0.382 | 0.379 | 0.339 | 0.369 | 0.372 | 0.326 | 0.343 | 0.388 |
| $\nu$ | 0.374 | 0.352 | 0.377 | 0.297 | 0.337 | 0.368 | 0.287 | 0.304 | 0.371 |
| $\zeta$ | 0.560 | 0.471 | 1.101 | 0.286 | 0.419 | 0.789 | 0.284 | 0.310 | 0.634 |
| A$_u$ | 2.926 | 2.821 | 0.114 | 3.051 | 2.788 | 0.396 | 2.577 | 2.909 | 1.663 |
| A$_k$ | 1.030 | 1.000 | 0.050 | 1.065 | 0.991 | 0.171 | 0.929 | 1.025 | 0.642 |
| V$_t$ | 2994.482 | 2261.203 | 1651.718 | 3265.373 | 2606.191 | 2142.493 | 3540.554 | 3142.391 | 2301.222 |
| V$_l$ | 6672.716 | 4726.336 | 3722.619 | 6076.384 | 5253.476 | 4684.118 | 6478.689 | 5916.928 | 5077.601 |
| V$_a$ | 3377.697 | 2542.844 | 1863.994 | 3645.956 | 2924.997 | 2414.631 | 3948.317 | 3948.317 | 2594.597 |
| $\theta_D$ | 490.796 | 352.350 | 241.320 | 538.772 | 414.970 | 323.163 | 591.467 | 507.397 | 355.902 |
| k | 3.632 | 3.036 | 3.746 | 2.129 | 2.730 | 3.447 | 2.015 | 2.212 | 3.535 |
| $H^{C_V}$ | 0.843 | 0.163 | 1.446 | 5.182 | 1.407 | 0.538 | 6.827 | 4.245 | 0.287 |
| $H^{T_V}$ | 2.907 | 2.184 | 0.980 | 6.249 | 3.167 | 1.671 | 7.682 | 5.452 | 1.893 |
| $\lambda$ | 119.488 | 47.786 | 28.077 | 73.717 | 59.343 | 47.111 | 83.224 | 68.270 | 60.382 |
| $\mu$ | 40.293 | 20.172 | 9.117 | 50.395 | 28.761 | 16.947 | 61.722 | 44.175 | 21.050 |

and the same has been used to compute the values as listed in Table 2. These numbers indicate that all compositions are inherently hard at different applied pressures. The elastic constants and moduli can also be used to calculate the average sound velocity. The compositions' transverse ($V_t$) and longitudinal ($V_l$) elastic wave velocities are also calculated using the relations listed as

$$V_l = \sqrt{\frac{3B + 4G}{\rho}}; V_t = \sqrt{\frac{G}{\rho}} \qquad (24)$$

where $\rho$ represents the mass density of the composition(Anupam. et al., 2024). The findings reported in Table 2 demonstrate that elastic constants do not increase as pressure increases, and the values of other parameters peak at 0 GPa, indicating that this pressure is essential for the compositions under study. Also, LiPdH$_3$ is also extremely resistive and incapable of experiencing volume changes, as evidenced by its greatest value of B at a pressure of 0 GPa. Additionally, we observe that LiPdH$_3$ has the maximum value of G at 10 GPa of pressure, indicating that it has a very rigid structure at this pressure. A material's Young's modulus (E), sometimes referred to as the modulus of elasticity, is a characteristic that indicates how stiff a solid is under applied stress. The more rigid the material, the higher the value of E. Table 2 shows that, for both compositions under study, the Young's modulus increases as pressure increases. Moreover, among the three compositions under study, LiPdH$_3$ is the stiffest composition at all examined pressures. LiPdH$_3$ has the greatest value of E at a pressure of 10 GPa, indicating that it is significantly stiffer than LiXH$_3$ (X=Ag,Cd). Poisson's

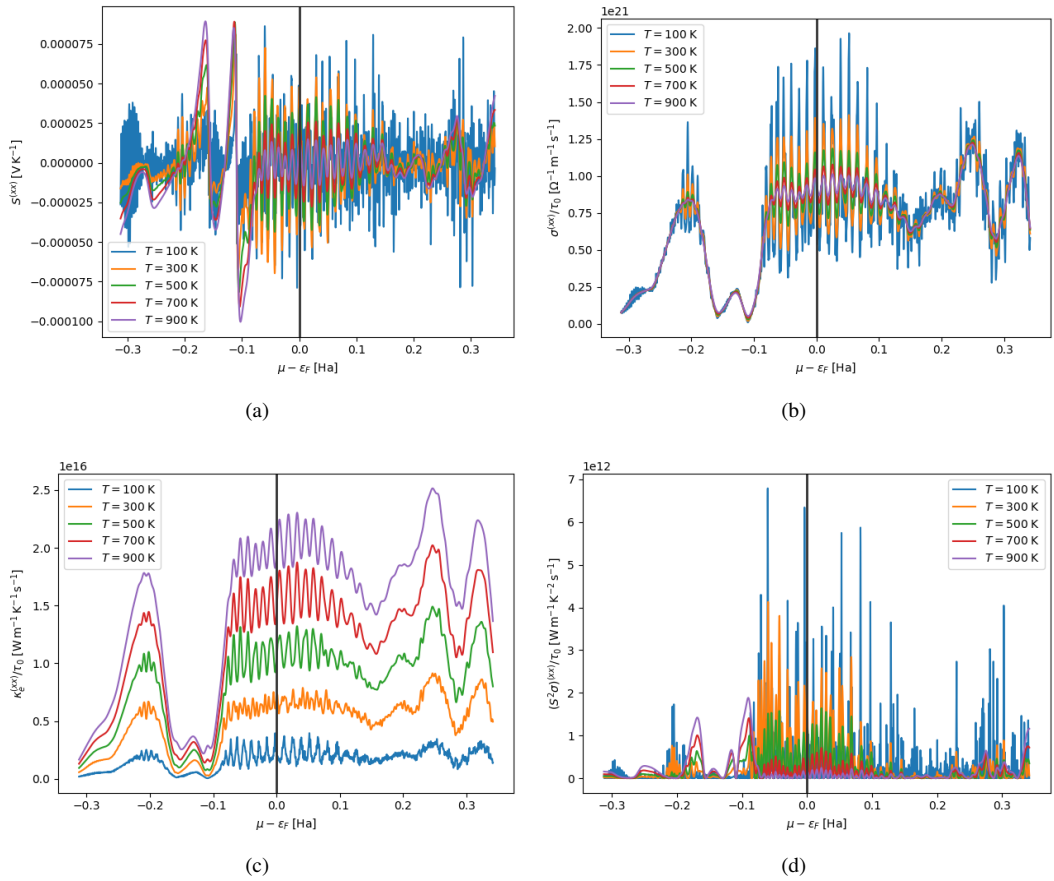

**Figure 6**: [a] Seebeck coefficient (S), [b] Electrical Conductivity ($\sigma$), [c] Electronic thermal conductivity ($\kappa_e$), [d] Power Factor ($S^2\sigma$) as a function of ($\mu - \epsilon_F$) at different temperatures for LiPdH$_3$

ratio provides information about a material's bonding properties. It indicates that a substance has an ionic character when its value is around 0.25 and covalent bonding when its value is around 0.1. Different Poisson's ratio values, as shown in Table 2, show that ionic bonding predominates in different compositions and that this dominance declines as pressure increases. Additionally, compared to LiXH$_3$ (X=Ag,Cd), LiPdH$_3$ has a stronger ionic nature. The $\frac{B}{G}$ ratio, represented as k, is used to calculate the brittleness or ductility of a material. The material is brittle if this ratio is less than 1.75; otherwise, it is ductile. The examined compositions are ductile in character, according to the listed values of k shown in Table 2. In addition, we have calculated the Kube-Anisotropy Index ($A_k$), Ranganathan-Anisotropy Index ($A_u$), Transverse elastic wave velocity ($V_t$ in m/s), Longitudinal elastic wave velocity ($V_l$ in m/s), Average wave velocity ($V_a$ in m/s), Debye Temperature ($\theta_D$ in K), Lame's first parameter ($\lambda$ in GPa), Lame's second parameter ($\mu$ in GPa) of LiXH$_3$ (X=Pd, Ag,Cd) perovskites under various external stress levels of 0, 5, and 10 GPa in order to account for stability of these compositions (Razzaq, S. et al., 2022). These numbers show that the compositions under study are stable at various pressures.

### 3.5. Thermoelectric Properties

In order to end the energy crisis and pave the way for sustainable energy solutions, waste heat may be converted into electrical energy using the effect known as thermoelectric effect(Ahmed, R., Masuri, N. S., Haq, B. U., Shaari, A., AlFaifi, S., Butt, F. K., Muhamad, M. N., Ahmed, M. and Tahir, S. A., 2017). This objective of converting thermal energy into electrical energy can be accomplished by thermoelectric materials. Low thermal conductivity is a characteristic of efficient thermoelectric materials. Therefore, maximizing S and $\sigma$ and minimizing the degree of heat conductivity $\kappa$ are required to have good thermoelectric material. We calculated the Seebeck coefficient (S), electrical conductivity ($\sigma$), electronic thermal conductivity ($\kappa_e$), power factor ($S^2\sigma$), and other thermoelectric properties as functions of ($\mu - \epsilon_F$) at various temperatures in order to investigate the thermal properties of lithium-based perovskites, LiXH$_3$ (X=Pd, Ag, and Cd). The results are plotted in Figures (6–8) for different parameters. The voltage that results from introducing a temperature gradient between two junctions is known as the Seebeck coefficient. Figures (6-8)a exhibit the Seebeck coefficient for compositions LiXH$_3$ (X=Pd,Ag,Cd). These figures clearly show that for (X=Pd), the Seebeck coefficient increases with temperature, while

LiXH$_3$ (X = Pd, Ag, Cd) as Hydrogen Storage Applications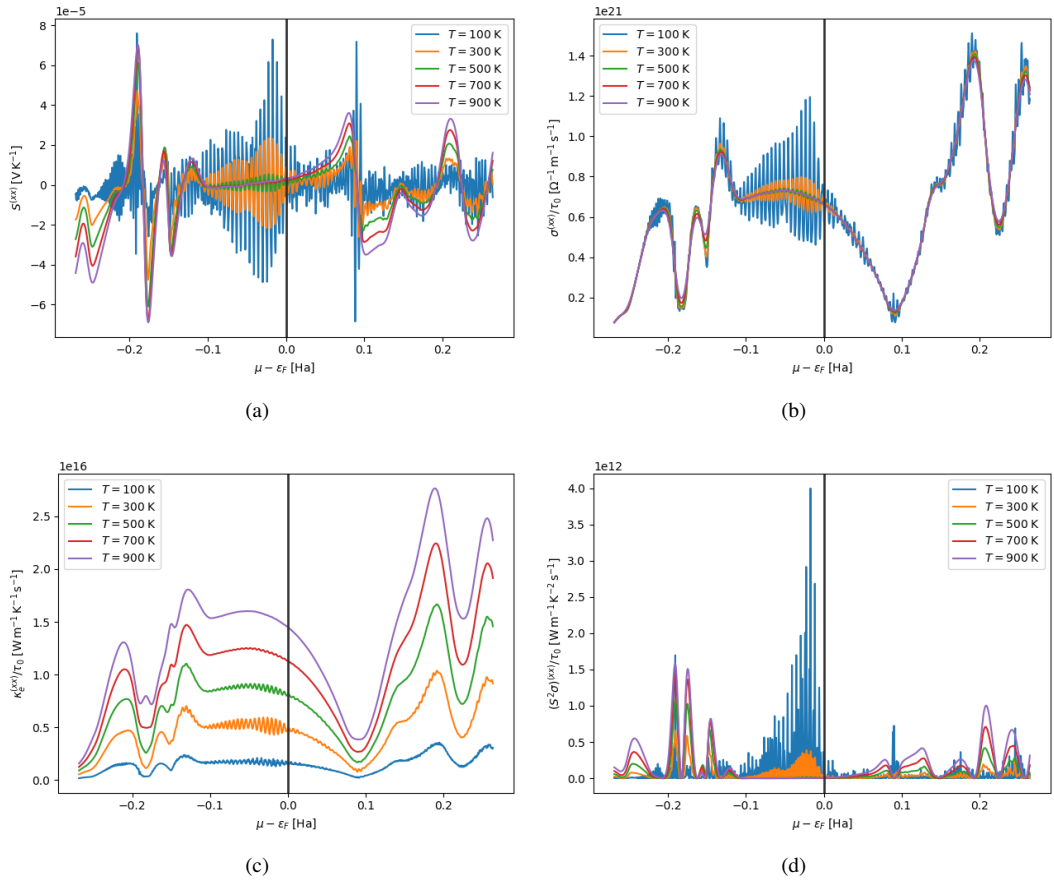

**Figure 7**: [a] Seebeck coefficient (S), [b] Electrical Conductivity ($\sigma$), [c] Electronic thermal conductivity ($\kappa_e$), [d] Power Factor ($S^2\sigma$) as a function of ($\mu - \epsilon_F$) at different temperatures for LiAgH$_3$

for (X=Ag, Cd), it decreases. The highest value of (S) for these respective compositions of LiXH$_3$ are ∼ 7.5 x 10$^{-5}$ VK$^{-1}$ at -0.17Hartree for 900 K for, ∼ 7.9 x 10$^{-5}$ VK$^{-1}$ at -0.19 Hartree for100 K and ∼ 1.5 x 10$^{-4}$ VK$^{-1}$ at 0.3 Hartree for 100 K. These observations indicate the dominance of phonon-phonon interaction in LiPdH$_3$ while that of either electron-phonon or electron-electron interactions in LiXH$_3$(X=Ag,Cd). When electrons go from hot areas to cold areas, current is generated. A prerequisite for materials with good thermoelectric characteristics is high electrical conductivity?. Figures (6-8)b illustrate the variation in electrical conductivity as a function of chemical potential for the compositions LiXH$_3$ (X=Pd,Ag, and Cd) in the temperature range of 100K to 900K. These figures demonstrate that the electrical conductivity values for LiPdH$_3$, LiAgH$_3$ and LiCdH$_3$ respectively, are $\sigma$ ∼1.79 x 10$^{21}$ $\Omega^{-1}m^{-1}s^{-1}$, ∼1.8 x 10$^{21}\Omega^{-1}m^{-1}s^{-1}$ and ∼1.5 x 10$^{21}\Omega^{-1}m^{-1}s^{-1}$ and concludes that as the temperature rises, the conductivity is observed to remain rather constant. The movement of electrons and lattice vibrations, or thermal conductivity, is also related to the transfer of heat within a material. Total thermal conductivity ($\kappa_e$) in a material is sum of electronic ($\kappa_{ele}$) and lattice thermal conductivity ($\kappa_{latt}$). Figures (6-8)c displays the thermal conductivity for the compositions LiXH$_3$ (X=Pd,Ag, and Cd) under study as a function of chemical potential throughout a range of temperatures. As per these plots a temperature results in increase of peak value of thermal conductivity. The thermal conductivity peak values for LiPdH$_3$, LiAgH$_3$ and LiCdH$_3$ at 900K are, respectively, $\kappa_e$∼2.5 x 10$^{16}$ $Wm^{-1}K^{-1}s^{-1}$, ∼2.8 x 10$^{16}Wm^{-1}K^{-1}s^{-1}$ and ∼2.4 x 10$^{16}Wm^{-1}K^{-1}s^{-1}$. These perovskites are therefore appropriate for thermoelectric applications involving high temperatures. The power factor (PF) of a thermoelectric device determines its partial efficiency. It may be expressed mathematically as $PF = \frac{S^2\sigma}{\tau}$.The (PF) for the compositions under study as a function of chemical potential at various temperature ranges is shown in Figures (6-8)d. From the plots, we noticed that the optimum value of (PF) as ∼6.9 x 10$^{12}Wm^{-1}K^{-2}s^{-1}$, ∼4 x 10$^{12}Wm^{-1}K^{-2}s^{-1}$ and 1.9 x 10$^{12}Wm^{-1}K^{-2}s^{-1}$, respectively, for LiXH$_3$ (X=Pd,Ag,Cd). Because of this, in addition to their ability to store hydrogen, these compositions exhibit significant thermoelectric response and can be used in thermoelectric devices as a substitute for other renewable energy sources. Zhang, J., Liu, H. J., Cheng, L., Wei, J., Lang, J. H., Fan, D. D., Shi, J., Tang, X. F. and Zhang, Q. J. (2014).

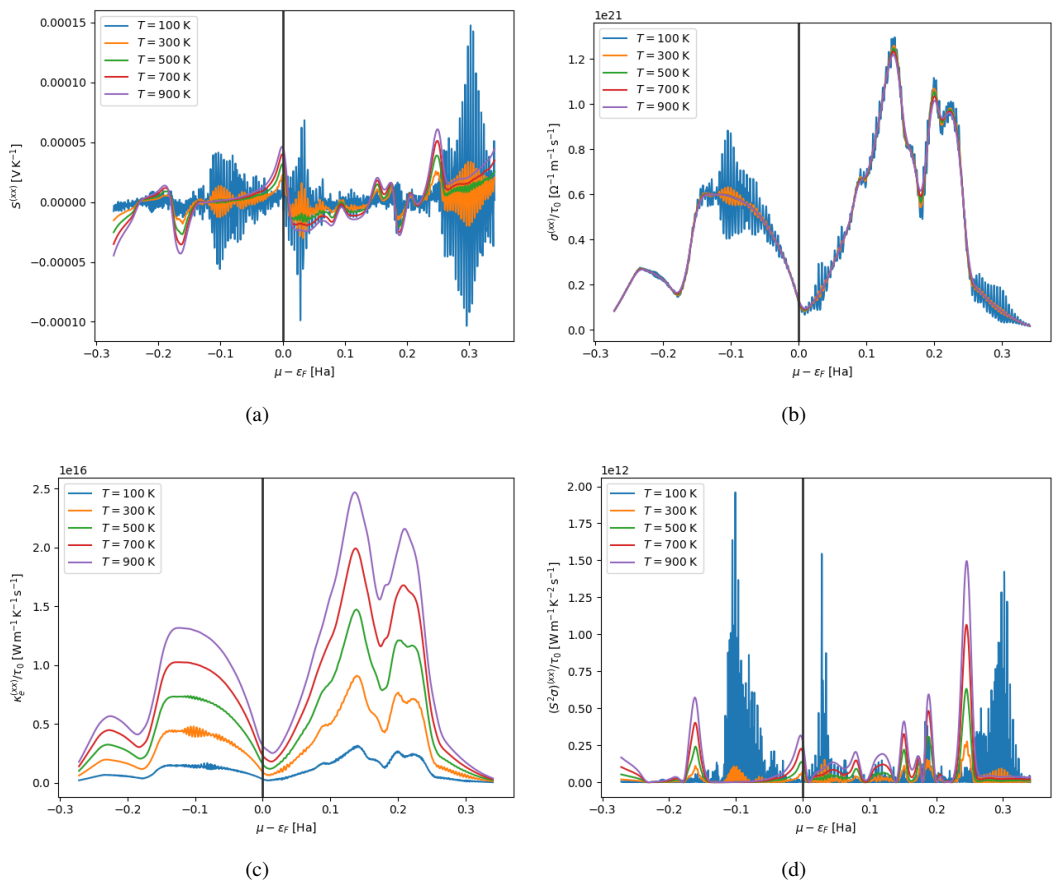

**Figure 8**: [a] Seebeck coefficient (S), [b] Electrical Conductivity ($\sigma$), [c] Electronic thermal conductivity ($\kappa_e$), [d] Power Factor ($S^2\sigma$) as a function of ($\mu - \epsilon_F$) at different temperatures for LiCdH$_3$

## 4. Conclusion

Using the DFT-based WIEN2k code, the structural, electrical, mechanical, and thermoelectric characteristics of the lithium-based perovskites LiXH$_3$ (X=Pd,Ag,Cd) are investigated. According to the structural investigations, all compositions in space group *Pm*3*m* [no. 221] have a stable cubic structure. They are metallic in nature, as shown by plots of the energy band gap and the density of states. Notably, the FP-LAPW approach gives us a enthalpy of formation as -55.05 kJ/mol.H$_2$, -24.36 kJ/mol.H$_2$ and -49.71 kJ/mol.H$_2$ for LiXH$_3$ (X=Pd,Ag,Cd). perovskites, which is substantially greater than the optimal value, which is ∼ - 40 kJ/mol.H$_2$, according to the US Department of Energy, indicating these perovskites except LiXH$_3$ (X=Ag) are thermodynamically stable. Furthermore, the enhanced hydrogen storage cyclability of these compositions is confirmed by the low decomposition temperatures. The compositions under investigation have promising hydrogen storage potential, as demonstrated by their respective gravimetric hydrogen storage capacities of 2.71%, 2.57%, and 2.47% for LiXH$_3$ (X=Pd,Ag,Cd). Furthermore, the mechanical stability of these materials is supported by the Born stability criterion. They also showed strong thermoelectric responses, which made them suitable materials for use in thermoelectric devices as alternative green energy sources in addition to their potential for storing hydrogen.

## 5. Declaration of Competing Interest

The authors have no conflict of interest for the work reported in this manuscript.

## CRediT authorship contribution statement

**Anupam:** Software, Workstation, Data generation. **Shyam Lal Gupta:** Conceptualization, Methodology, Data curation, Writing - original draft Writing - review and editing. **Sumit Kumar:** Software, Workstation, Data generation. **Samjeet Singh Thakur:** Data compilation, Electronic structure. **Sanjay Panwar:** Software, Workstation, Data generation. **Diwaker:** Conceptualization, Methodology, Data curation, Writing - original draft Writing - review and editing.